\documentclass[twocolumn,a4paper,superscriptaddress,floatfix,showpacs]{revtex4}
\usepackage{amsmath,amssymb,epsfig,color,textcase}
\usepackage{natbib}

\begin{document}

\title{Orbital-dependent singlet dimers and orbital-selective
Peierls transitions in transition metal compounds}

\author{Sergey V. Streltsov}
\affiliation{Institute of Metal Physics, S.Kovalevskoy St. 18, 620990 Ekaterinburg, Russia}
\affiliation{Ural Federal University, Mira St. 19, 620002 Ekaterinburg, Russia}
\email{streltsov@imp.uran.ru}

\author{Daniel I. Khomskii}
\affiliation{II. Physikalisches Institut, Universit$\ddot a$t zu K$\ddot o$ln,
Z$\ddot u$lpicher Stra$\ss$e 77, D-50937 K$\ddot o$ln, Germany}


\date{\today}

\begin{abstract}
We show that in transition metal compounds containing structural metal dimers there may exist in 
the presence of different orbitals a special state with partial formation of singlets by 
electrons on one orbital, while others are effectively decoupled and may give e.g. long-range magnetic 
order or stay paramagnetic. Similar situation can be realized in dimers spontaneously formed at structural 
phase transitions, which can be called orbital-selective Peierls transition.  This can occur in 
case of strongly nonuniform hopping integrals for different orbitals and small intra-atomic Hund's rule 
coupling $J_H$.  Yet another consequence of this picture is that 
for odd number of electrons per dimer there exist 
competition between double exchange mechanism of ferromagnetism, and the formation of singlet dimer 
by electron on one orbital, with remaining electrons giving a net spin of a dimer. 
The first case is realized for strong Hund's rule coupling, typical for $3d$ compounds, whereas the second is 
more plausible for $4d-5d$ compounds.
We discuss some implications of these phenomena, and consider examples of real 
systems, in which orbital-selective phase seems to be realized.
\end{abstract}

\maketitle

{\it Introduction.--} 
Molecular-like clusters exist in many inorganic transition metal (TM) compounds. Sometimes these 
are determined just by the crystal structure, like e.g. dimers in
CuTe$_2$O$_5$~\cite{Hanke1973,Ushakov2009} or trimers in 
Ba$_4$Ru$_3$O$_{10}$~\cite{Streltsov2012a}. However such molecular clusters 
may also appear spontaneously from a homogeneous solid, e.g. due to 
Peierls or spin-Peierls transition, which results in the
formation of dimers in VO$_2$~\cite{Imada1998},
MgTi$_2$O$_4$,~\cite{Khomskii2005a} or CuGeO$_3$~\cite{Hase1993}, trimers in LiVO$_2$,~\cite{Pen1997} 
tetramers CaV$_4$O$_9$,~\cite{Starykh1996} 
or even heptamers like in AlV$_2$O$_4$~\cite{Horibe2006}.
In many such cases the TM ions have several electrons in different orbitals state, and 
often just one particular orbital is responsible for the formation of a molecular cluster. 
The question arises, what is in such a case the role and the ``fate'' of other electrons 
which can exist on a TM ion.

Usually the intra-atomic Hund's rule exchange $J_H$  binds all electrons of an ion into 
a state with maximum spin, and,
e.g., when one particular electron on a certain orbital forms a valence bond with the 
neighboring site, other electrons just follow, so that all 
electrons are in a spin singlet state with the neighboring site. However it is not the 
only possibility. One can argue that if the intersite electron hopping 
is large compared with the Hund's exchange (which can happen especially in 
$4d$ and $5d$ systems, in which the covalency is strong, but $J_H$ is 
reduced), only one ``active'' electron at a site would participate in the formation of 
a molecular orbital (MO), the other electrons being, in a sense, decoupled and can live their own 
life. For example if the remaining electrons would interact with other sites, they can form 
some magnetically-ordered state, which would coexist with the molecular orbitals formed by 
``active'' electrons. That is, in this case we can speak about 
the {\it orbital-dependent dimer formation}, or  {\it orbital-dependent Peierls transition.}
The same mechanism leads to the competition between double exchange ferromagnetism and the formation 
of singlet dimers for fractional number of electrons per center. 
\begin{figure}[b]
 \centering
 \includegraphics[clip=false,width=0.35\textwidth]{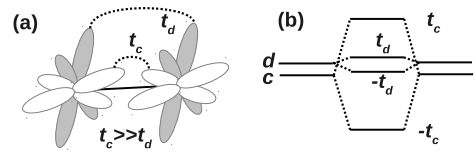}
\caption{\label{levels}(a) The sketch illustrating how different
hoppings integrals may appear in the system. The $t_{dd\pi}$ hopping 
between white ($t_{2g}$) orbitals is larger than between grey 
$t_{dd\delta}$. (b) Corresponding levels splitting.}
\end{figure}

In this Letter we substantiate this picture by different means, using analytical and
numerical calculations, and discuss some experimental examples, in which this 
phenomenon seems to take place.

{\it Model treatment.--}
Suppose we have a dimer, or a chain of dimers with different orbitals at each site, 
only one of which has a strong overlap with neighboring sites (white orbitals in 
Fig.~\ref{levels}a). Those orbitals provide strong intersite hopping $t_c$. If there 
will be two electrons per site in a dimer, then one electron is localized on the bonding combination of 
$c-$orbitals, while 
another electron can occupy the orbital which has no, or much smaller
overlap and hopping $t_d$ with the neighbors (shaded orbitals in Fig.~\ref{levels}a). 
These localized electrons ($d-$electrons) interact with the ``mobile'' $c-$electrons by 
the Hund's rule interaction
$H_{Hund} = -J_H (\frac 12 + 2 \vec S_{id} \vec S_{ic})$,
where $S_{id}$ and $S_{ic}$ are spins of localized and mobile 
electrons at site $i$. And of course all electrons in 
principle would experience a local (Hubbard) repulsion $U$. Thus the total model 
for the case of two different orbitals per site can be written in following form:
\begin{eqnarray}
H &=& \sum_{<ij>\sigma} (- t_c c^{\dagger}_{i \sigma}c_{j \sigma} - t_d 
d^{\dagger}_{i \sigma}d_{j \sigma}) + U \sum_{i \sigma \sigma'} n_{i \sigma} n_{i \sigma'}
\nonumber \\
&-&J_H \sum_i (\frac 12 + 2 \vec S_{id} \vec S_{ic}) +H_{inter}.
\label{Ham}
\end{eqnarray}
Here the first three terms describe electrons in a dimer, and 
$H_{inter}$ takes into account electron hopping and other coupling terms 
(e.g. the exchange interaction) between dimers. The ground state of a 
dimer would be different depending on 
the ratios of different parameters in Eq.~\eqref{Ham}.

1) If we first ignore Hubbard interaction and assume the strong  
hopping, $t_c \gg J_H$  (and $t_d$ is small), the $c-$electrons would form a 
singlet state, a bonding state described by usual MO 
(see Fig.~\ref{levels}b):
$|MO \rangle = \frac 12 |(c^{\dagger}_{1 \uparrow} + c^{\dagger}_{2 \uparrow})
( c^{\dagger}_{1 \downarrow} + c^{\dagger}_{2 \downarrow}) \rangle.$
The other electrons, not participating in formation of this singlet, would be 
then effectively decoupled, i.e. the total wave function would be
\begin{equation}
\label{MOtilda}
|\widetilde {MO} \rangle = \frac 1 2 |(c^{\dagger}_{1 \uparrow} + c^{\dagger}_{2 \uparrow})
                       (c^{\dagger}_{1 \downarrow} + c^{\dagger}_{2 \downarrow}) \Psi_d \rangle. 
\end{equation}
The $d-$electrons can be also described by the singlet state (but in Heitler-London (HL) 
form more appropriate for localized electrons): $\Psi_d^{HL} = |d^{\dagger}_{1\uparrow}  
d^{\dagger}_{2\downarrow}  -d^{\dagger}_{1\downarrow}  d^{\dagger}_{2\uparrow} \rangle /\sqrt 2$, or
by other combinations of localized spins, e.g. 
$\Psi_d^{AFM} = |d^{\dagger}_{1\uparrow}  d^{\dagger}_{2\downarrow} \rangle$
or $\Psi_d^{FM} = |d^{\dagger}_{1\uparrow}  d^{\dagger}_{2\uparrow} \rangle$
to model partially ordered states. A particular choice of $\Psi_d$ depends 
on properties of system, but the orbital-selective behavior can
be observed in any of them. If we chose the HL form of 
$\Psi_d^{HL}$, then one gains in $\widetilde {MO}$ the full bonding energy given by $t_c$, but do 
not lower the total energy due to the Hund's term, 
$\langle \widetilde {MO}|H_{Hund}|\widetilde {MO} \rangle$=0. 

2) If, instead, $J_H>t_c(\gg t_d)$, then first of all the strong Hund's 
exchange would couple two spins at a site into one common state with $S=1$, and then we 
should form a singlet out of these two states $S=1$ at neighboring sites. Corresponding wave 
function would have the form~\cite{LandauQ}
\begin{eqnarray}
|\widetilde {HL} \rangle &=& | S_{tot}=0 \rangle = \frac 1{\sqrt 3}
( |S^z_1=1, S^z_2=-1 \rangle \nonumber\\  
&+& |S^z_1=-1, S^z_2=1 \rangle - |S^z_1=0, S^z_2=0 \rangle)\nonumber \\ 
&=& \frac 1 {\sqrt 3} \Big(| c^{\dagger}_{1 \uparrow} d^{\dagger}_{1 \uparrow} 
                        c^{\dagger}_{2 \downarrow} d^{\dagger}_{2 \downarrow}\rangle +
                      | c^{\dagger}_{1 \downarrow} d^{\dagger}_{1 \downarrow} 
                        c^{\dagger}_{2 \uparrow}   d^{\dagger}_{2 \uparrow}\rangle \nonumber \\
&-&
             \frac 12 | 
                         (c^{\dagger}_{1 \uparrow}   d^{\dagger}_{1 \downarrow} 
                        + c^{\dagger}_{1 \downarrow} d^{\dagger}_{1 \uparrow})
                         (c^{\dagger}_{2 \uparrow}   d^{\dagger}_{2 \downarrow} 
                        + c^{\dagger}_{2 \downarrow} d^{\dagger}_{2 \uparrow})
                         \rangle \Big).
\label{HL}
\end{eqnarray}
We see that for strong Hund's coupling 
the dimer wave function has actually not MO, but HL form, the two-electron analogue 
of the usual HL wave function 
$|HL \rangle = \frac 1{\sqrt 2}|c^{\dagger}_{1\uparrow} c^{\dagger}_{2 \downarrow} - 
        c^{\dagger}_{1 \downarrow} c^{\dagger}_{2 \uparrow} \rangle$:
it does not contain ionic configurations of the type $c^{\dagger}_{1\uparrow} c^{\dagger}_{1 \downarrow}$, etc.
For this state we gain the full Hund's energy, but lose large part of the bonding energy,
which for $J_H >(t_c,t_d)$ is more favorable.

 The same state,~\eqref{HL}, one would get also when we have strong Hubbard interaction 
[$(U, J_H) >(t_c,t_d)$]. It is interesting to notice that both, strong Hund's exchange and Hubbard 
repulsion lead to localization of electrons 
at respective sites, and to the HL wave function. This is 
reminiscent of the notion of Hund's metal (or here rather Hund's insulators).~\cite{Haule2009,Georges2013}

These limiting cases 1) and 2) may be not very realistic, and one has to consider
intermediate values of parameters and include both the Hund's 
rule exchange $J_H$ and the Hubbard $U$. But we will see that the effect illustrated on the 
limiting case $J_H = U = 0$ -- the formation of a singlet state by electrons on one orbital, 
other electrons remaining decoupled and ``magnetic'', 
survive also in a more realistic case. 
For spontaneous dimerization, such as at a Peierls transition, this 
would mean that we have (strong coupling) orbital-selective Peierls transition.

In the general case we can consider this situation using variational procedure, taking the 
wave function as the superposition of the $\widetilde {MO}$ 
and $\widetilde {HL}$ states
\begin{equation}
\label{WF}
|\Psi \rangle = c(|\widetilde {MO} \rangle + \alpha |\widetilde {HL} \rangle),
\end{equation}
(where $c$ is the normalization factor) and minimizing total energy 
$\langle \Psi | H | \Psi \rangle$. 
For the simplisity the Hund's rule interaction will be treated 
in the mean field way, i.e. substituting $\vec S$ by $S^z$ in Eq.~\eqref{Ham}. 
Straightforward calculations show that indeed the solution approaches to pure $\widetilde {MO}$ 
state for $t_c \gg (U, J_H)$, and tends to the $\widetilde {HL}$ state in the opposite 
limit. For $J_H  \ll t_c$  the  coefficient $\alpha \sim J_H/t_c$. 
In the opposite 
limit $J_H \gg t_c$ the solution tends to the pure HL state, $1/\alpha \sim t_c/J_H$.
For intermediate values the system gradually switches from one regime, in which the first 
electron forms singlet dimer with the neighbor, and the second is largely 
decoupled  (orbital-selective dimer formation), to a state in which both electrons are 
in a singlet state.

Using the wave function \eqref{WF} with the coefficient $\alpha$ determined variationally, we can also 
find the value of the average spin at a site, e.g. for the case of antiparallel orientation of 
spins at two sites, i.e. taking 
$\Psi_d^{AFM} = |d^{\dagger}_{1\uparrow}  d^{\dagger}_{2\downarrow} \rangle$ in \eqref{MOtilda}, and 
corresponding part of the HL wave function  (the first term in \eqref{HL}, with proper normalization). The 
coefficient $\alpha$ in this 
case is $\alpha = J_H/2t_c$ for $J_H/t_c \ll 1$, and $\alpha = 2J_H/t_c$ in the opposite limit. Average spin 
$\langle S^z_i \rangle$ on the $i$th site in this case interpolates between the values 
$\langle S^z_i \rangle =1/2$ for $J_H=0$ and the ``full'' value $\langle S^z_i \rangle =1$ for very large $J_H$. 
Asymptotic behavior is  $\langle S^z_i \rangle =1/2+J_H/4t_c$ for $J_H/t_c \ll 1$, and 
 $\langle S^z_i \rangle =1-(t_c/J_H)^2/8$ for $J_H/t_c \gg 1$.

It is important to note that at intermediate values of $J_H/t_c$ the average spin at a site has the 
value intermediate between 1/2 and 1, i.e. the magnetic moment is $1\mu_B < \mu < 
2\mu_B$. It is this moment, strongly reduced as compared with  $2\mu_B$ usually expected for $d^2$ 
configuration, which would be seen in susceptibility and which could eventually participate in magnetic 
ordering. Such strong quenching of  a moment in such systems may be a signature of partial 
orbital-selective dimerization.

We can also take into account Hubbard repulsion between electrons in variational
procedure. The results are very similar to the case of Hund's coupling only, with the substitution 
$J_H \to J_H +U/4$ (for the spin-ordered state considered above). 
Thus we see that if both $(J_H, U) \ll t_c$, the system is in an 
orbital-selective regime ($c-$electrons form singlet dimers, $d-$electrons are effectively
decoupled from those); and for $(J_H, U) \gg t$ (either both, or at least one of them), we have a 
HL state~\eqref{HL} with the total spin $S$ per site 
and suppressed ionic configurations. We see that the strong Hubbard and 
Hund's couplings 
act in the same direction: they both suppress MO state, localize electrons at particular sites and couple spins at 
the same ion into a total spin $S$. For strong Hubbard interactions 
$U \gg t$ already a relatively 
weak Hund's coupling $J_H > t^2/U$ is sufficient for that. But in principle we can get HL state 
only due to the strong Hund's coupling, even without Hubbard repulsion.

\begin{figure}[t!]
 \centering
 \includegraphics[clip=false,width=0.35\textwidth,angle=270]{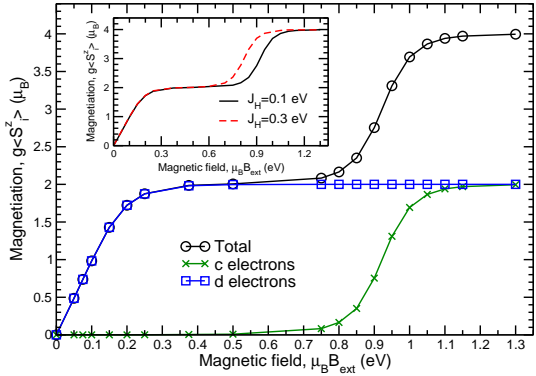}
\caption{\label{Field}(color online). 
The total and partial magnetization per dimer, calculated in C-DMFT. 
$t'=0.1$ eV, $t_d=0.2$ eV,
$t_c=6t_d$, $J_H=t_d/2$, $U=5 t_d$, $T=0.1$ eV. Inset shows
dependence of total magnetization on Hund's rule exchange.
}
\end{figure}

{\it DMFT calculations.--}  To check the treatment presented above we consider a model system 
- one dimensional chain of dimers, using the cluster 
extension of the dynamical mean field theory (C-DMFT)~\cite{Biroli2002}
with Hirsh-Fye (HF-QMC) solver.~\cite{Hirsch1986}
There are two orbitals and two electrons per site in the dimer.
Intradimer hoppings are $t_d$ and $t_c$, interdimer $-t'$ is the same for both orbitals 
and allowed only for the neighboring
sites. We neglected the intersite Coulomb interaction, 
so that
the sites are coupled by the kinetic energy term only. The on-site Coulomb
repulsion term was taken to be
$U_{m m}^{\sigma \sigma'} = U$, $U_{m m'}^{\sigma \sigma'} = U-2J_H$, 
$U_{m m'}^{\sigma  \sigma} = U - 3J_H$.
The Hund's rule exchange was considered in the Ising form.

The field dependence of the magnetization presented in Fig.~\ref{Field} shows
that there is no magnetic response in a zero external field
(as here both $t_c$ and $t_d$ are nonzero, the ground state of a dimer is a 
singlet for both electrons).
An increase of $B_{ext}$
drives the systems to the orbital-selective regime, when $c-$electrons initially are predominantly in 
the MO singlet state, while $d-$electrons are detached and start to be polarized 
only at higher fields also $c-$electron singlet is broken and $c-$electrons become polarized.
 As it was argued above an internal exchange field (e.g. Heisenberg exchange) may result in a similar 
situation. Moreover the range of the orbital-selective phase depends on the $J_H/t_c$ ratio 
(see inset of Fig.~\ref{Field}).

\begin{figure}[t!]
 \centering
 \includegraphics[clip=false,width=0.35\textwidth,angle=270]{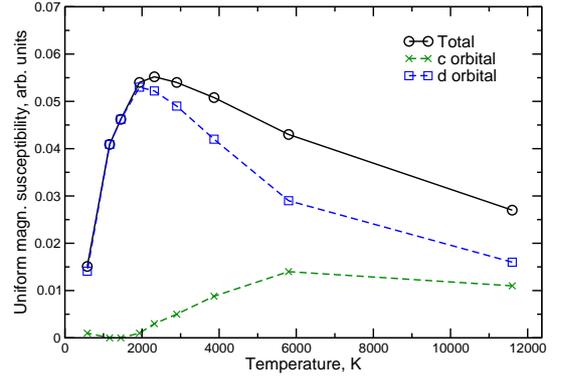}
\caption{\label{chi}(color online).
Uniform magnetic susceptibility, calculated in C-DMFT as $\chi = M/B_{ext}$,
where $M$ is magnetization per dimer, and $B_{ext}$ external magnetic
field. $t'=0.1$ eV,  $t_d=0.4$ eV, $B_{ext}=0.1$ eV,
$U=5.25t'$, $t_c=3t_d$, $J_H=1.25t_d$.}
\end{figure}

Different character of the orbitals is also reflected in the
temperature dependence of the uniform magnetic susceptibility $\chi(T)$.
It is seen in Fig.~\ref{chi} that overall temperature behavior of $\chi$ 
is consistent with what one may expected for dimers: drastic 
decrease at low temperatures (LT) due to the spin singlet state formation 
and Curie-like tail at high temperatures.
However partial contributions to the susceptibility is again quite
different.  Orbital with the smallest hopping provides the largest contribution
at low T. Corresponding electrons behave as free spins at intermediate
temperatures, whereas $c-$electrons are still in a singlet dimer state.
Only with further increase of the temperature the second orbital starts to contribute. 
This may result in the shift 
of the magnetic susceptibility maximum and has to be taken into
account in the fitting procedures (to evaluate exchange integrals)
for systems with the orbital-selective behavior.

Thus these results indeed confirm our model treatment presented above: for chosen 
parameters one may observe formation of the orbital-selective singlet state, which, 
if we start from a regular system and make spontaneous dimerization, would correspond 
to orbital-selective Peierls transition.
 
{\it Real materials.--}  As we saw above, orbital-selective singlet state can occur 
for specific conditions: when hopping for one orbital in a dimer is comparable or larger than 
the intra-atomic Hund's exchange (and Hubbard repulsion). This is less likely in 
$3d$ systems, for which $U$ or $J_H$ are usually larger than hopping ($U\sim$3-6 eV,
$J_H\sim$0.7-1.0 eV), and this is why this situation is not realized in V$_2$O$_3$,~\cite{Ezhov1999} 
as it was proposed by Castellani et al.~\cite{Castellani1978a}

But such state could easily appear in $4d$ and $5d$ systems, where both $J_H$ and $U$ are 
strongly reduced, while $t$ is getting larger. Thus for $5d$ metals
typically $U\sim$1-2 eV, $J_H\sim$0.5 eV, but radius of $5d-$orbitals is larger than of 
$3d$, and we can get to the situation with $dd$ hopping at least of order, or larger than 
($U$, $J_H$).

Such a situation may be met in some systems with dimerization, e.g. Li$_2$RuO$_3$,
where Ru-Ru dimers are formed in the common edge (of RuO$_6$ octahedra) geometry. 
The hopping between two $xy$ orbitals directed to each other in the dimer is $\sim$1.2 eV, which is much larger than between any other of $t_{2g}$ 
orbitals ($\sim$0.3 eV).~\cite{Kimber2013} This may explain why in the high 
temperature phase magnetic susceptibility behaves as for paramagnetic 
$S=1/2$, not $S=1$ centers (as it should be for Ru$^{4+}$).~\cite{Kimber2013}

Also some $3d$ compounds can show the behavior described above, although it is 
less likely than for $4d$ and $5d$ systems. Most probably this is the situation 
in V$_4$O$_7$.~\cite{Botana2011, Heidemann1976, Gossard1974}. The NMR data suggest
that there is  a partial formation of singlets in V$^{3+}$ chains, with the remaining 
magnetic moment of V$^{3+}$ ($d^2$) strongly reduced.~\cite{Heidemann1976} Thus, though 
this system is hardly an example of complete decoupling of two electrons on each V, it 
is apparently ``half-way'' to this regime.

Yet another realization of orbital-selective dimerization can exist when 
electrons on one orbital form dimers, but 
the others fill three-dimensional bands, so that the resulting state 
could be a metal, but with dimers. Such state seems to exist in MoO$_2$.~\cite{Eyert2000}
MoO$_2$ has a rutile structure, and Mo ions form dimers similar to those existing below the famous 
metal-insulator transition in VO$_2$. But, whereas in VO$_2$ there is one electron 
per site, which form singlet dimers, so that the LT state of VO$_2$ is a 
diamagnetic insulator, in Mo there are two electrons per Mo, one of which gives in 
MoO$_2$ the same dimers as in VO$_2$, and the other electrons 
provide metallic conductivity.
\begin{figure}[t]
 \centering
 \includegraphics[clip=false,width=0.5\textwidth]{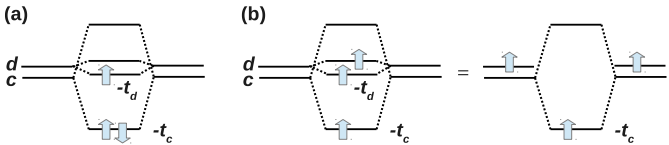}
\caption{\label{DE}
(color online). Possible orbital states in the case of the dimer with 2 orbitals 
per site and 3 electrons per dimer.}
\end{figure}

Special situation can exist if there is fractional occupation of $d-$levels, giving odd 
number, e.g. three, electrons per dimer~\cite{Streltsov2014b}. When $J_H>t$, one expects the usual double 
exchange (DE), which gives the state with all spins parallel (Fig.~\ref{DE}b) 
with the energy $E_{2b}=-t_c - J_H$, if the Hund's rule term in Eq.~\eqref{Ham}
is treated in a mean-field way.
In opposite case (Fig.~\ref{DE}a) two electrons form singlet bonding state, 
with remaining unpaired spin 1/2 per dimer and with the energy $E_{2a}=-2t_c-t_d - J_H/2$. 
Thus the DE ferromagnetism is realized if $J_H>2(t_c+t_d)$; 
in the opposite limit partial singlet formed on strongly overlapping orbitals suppresses DE and 
reduces total spin. The first situation is typically realized in $3d$ systems with large $J_H$, e.g. in 
Zener polarons in doped manganites
(note that Zener suggested this concept just for Mn dimers)~\cite{Zener1951,Daoud-Aladine2002}.
The alternative state, with partial 
singlets, is more plausible for $4d-5d$ systems, e.g. it was found 
in  Y$_5$Re$_2$O$_{12}$~\cite{Chi2003}.

{\it Conclusions.--}
Using analytical and numerical calculations we demonstrate that in systems with 
orbital degrees of freedom there may exist structural dimers
in the orbital-selective singlet state, or 
there may appear a (strong coupling)
orbital-selective Peierls transition: electrons on one orbital, 
having strong overlap and large hopping within the dimers, form a singlet state 
(bonding MO), whereas other electrons remain essentially decoupled and can, 
for example, give long range magnetic ordering (with strongly 
reduced moment) or stay paramagnetic. This situation resembles somewhat 
that of orbital-selective Mott 
transition.~\cite{Anisimov2002} 
For partial filling of $d-$levels, e.g. three electrons per dimer, this can 
lead to the suppression of the usual double exchange mechanism of ferromagnetism: 
mobile electrons can form singlets, the remaining electrons being decoupled from 
those. Typically such phenomena may occur when  
hopping between particular orbitals becomes larger than (or at least comparable to) the 
Hubbard repulsion $U$ and 
Hund's exchange $J_H$. It is not very plausible for $3d$ systems (although there 
are such examples); but it is likely for $4d-5d$ compounds, 
for which both $U$ and $J_H$ are strongly reduced, but the covalency and hopping 
are increased. We discuss different possible states which may appear in this situation, 
consider its possible experimental manifestation, and present some real examples of 
systems in which this physics seems to play a role.

{\it Acknowledgments.--}
We are grateful to A. Poteryaev, H. Jeschke, R. Valent\'i, I. Mazin for useful discussions.
This work is supported by the Russian Foundation for Basic Research 
via RFFI 13-02-00374, by the Ministry of  education and science of Russia 
through the program MK-3443.2013.2, by the German project 
FOR 1346 and by Cologne University via German excellence initiative.

\bibliography{../library}
\end{document}